\documentclass[pre,10pt,amsmath,amssymb]{revtex4}
\usepackage{graphicx}

\usepackage{amsmath,amsfonts,amssymb,bm}

\begin{document}

\draft
\title{Fluctuations in multiplicative systems with jumps}

\author
{Tomasz Srokowski}

\affiliation{
 Institute of Nuclear Physics, Polish Academy of Sciences, PL -- 31-342
Krak\'ow,
Poland }

\date{\today}

\begin{abstract}
Fluctuation properties of the Langevin equation including a multiplicative, 
power-law noise and a quadratic potential are discussed. The noise has 
the L\'evy stable distribution. If this distribution is truncated, the covariance 
can be derived in the limit of large time; it falls exponentially. 
Covariance in the stable case, studied for the Cauchy distribution, exhibits 
a weakly stretched exponential shape and can be approximated by the simple 
exponential. The dependence of that function on system parameters is determined. 
Then we consider a dynamics which involves the above process and obey 
the generalised Langevin equation, the same as for Gaussian case. The resulting 
distributions possess power-law tails -- which fall similarly to those for 
the driving noise -- whereas central parts can assume the Gaussian shape. 
Moreover, a process with the covariance $1/t$ at large time is constructed 
and the corresponding dynamical equation is solved. 
Diffusion properties of systems for both covariances are discussed. 
\end{abstract} 


\maketitle


\section{Introduction}

Trajectories encountered in complex systems often reveal discontinuities 
and the probability distributions are not governed by equations with local 
operators. Those distributions differ from the Gaussian and contain slowly falling 
tails. L\'evy stable distributions with tails $\sim |x|^{-1-\alpha}$, 
where $\alpha\in(0,2)$, are distinguished due to the generalised central 
limit theorem. However, the divergent variance makes the L\'evy stable distributions 
problematic in some physical applications; it may imply, for example, infinite 
kinetic energy. Similarly, covariance functions for the L\'evy stable processes 
with $\alpha<2$ do not exist. The above difficulties do not emerge if the tails, 
being still of the power form, fall faster than for the L\'evy stable distributions. 
In fact, tails of the form $|x|^{-\beta}$, where $\beta\ge3$, are frequently observed. 
This is the case for the financial market that possesses typical characteristics of 
the complex system and then some of its properties may be universal. 
Analyses of returns of stock indices show that a cumulative 
distribution of returns is power-law with $3<\beta<4$ \cite{sta,ple} and such values 
of $\beta$ are required by the optimal market strategy \cite{gab}. The minority game 
implies a similar value, $\beta=3.9$ \cite{ren}. Since large jumps represent extreme 
events, one can expect that first passage time probability should obey the Weibull 
distribution. However, a phenomenological analysis of the empirical data demonstrate 
that it is the case only for small returns, the large ones are of the power form 
with $\beta=3.32$ \cite{per}. Moreover, fast falling power-law tails result 
from a multifractal analysis of the extreme events \cite{muz}, characterize the hydraulic 
conductivity in the porous media and the atmospheric turbulence \cite{sche}.

The variance becomes finite when we modify the asymptotics of the L\'evy stable 
distribution by introducing either a simple cut-off or some fast-falling tail. 
Such truncated distributions very slowly \cite{man} converge with time to the normal distribution. 
Moreover, dynamical systems stimulated by the L\'evy stable noise possess finite moments 
of the stationary distribution if the particle is trapped inside a potential well with 
a sufficiently large slope \cite{che}. The Langevin equation with a multiplicative 
L\'evy noise $\eta$ and a linear deterministic force also predicts finite moments, 
if interpreted in the Stratonovich sense; this property was demonstrated by numerical 
simulations \cite{sro1,sro2}. On the other hand, we can define the coloured noise 
$\eta_c(t)$ by the generalised Ornstein-Uhlenbeck process, 
\begin{equation}
\label{wnoi}
d\eta_c(t)=-\gamma_n\eta_c(t)dt+\gamma_ndL(t),  
\end{equation}
where the increments of $L(t)$ have the stable L\'evy distribution, and take 
the white-noise limit, $\eta(t)=\lim_{\gamma_n\to\infty}\eta_c(t)$. 
Stochastic equation with the white noise $\eta(t)$, given by the above expression, 
allows us to change the variable in the usual way and obtain the Stratonovich 
result \cite{sron}. The generalised 
Wiener process with that noise as a driving force is characterised by the subdiffusive 
motion. Taking into account that the mass in the Langevin equation is finite modifies 
the slope of the tails: it diminishes with the inertia and finally converges to the result 
of the It\^o interpretation for the infinite mass. Multiplicative non-Gaussian 
white noises serve to describe population and ecological problems: a dynamics of two 
competing species \cite{dub1,dub2} and a population density in terms of 
the Verhulst model \cite{dub3,dub4}. 

Long tails of the distributions in complex systems are often accompanied by a long 
memory \cite{ren,seg} and then time-dependence of the fluctuations becomes non-trivial. 
The well-known multifractal structure of the financial time series \cite{osw} 
is attributed to both fat tails, that fall as a power-law 
but faster than for the stable L\'evy, and power-law correlations \cite{kant}. 
A slow decay of the correlations, even slower than a power-law, is regarded as a necessary 
condition of the multifractality in any complex system \cite{sai}. 
In the present paper, we demonstrate that the generalised Ornstein-Uhlenbeck process $\xi(t)$, 
driven by $\eta(t)$, possesses a well-determined covariance and fat tails 
of the distribution. Moreover, trajectories $\xi(t)$ preserve a typical feature 
of the L\'evy flights: smooth segments interrupted by large, rare jumps. 

In the presence of the memory effects, a description in terms of the standard Langevin 
equation with a coloured noise is problematic since a response of the system to the stochastic 
stimulation is not instantaneous and, as a consequence, a retarded friction must be introduced. 
The fluctuation-dissipation relation requires that the equipartition energy rule is satisfied, 
i.e. the temperature is well defined, and a retarded friction kernel is uniquely determined by the 
noise covariance; then $\xi(t)$ is called the internal noise. 
Obviously, that relation does not hold for the L\'evy noise $L(t)$ due 
to the infinite covariance \cite{west} and $L(t)$ can be regarded only as an external noise. 
The dynamical equation with the retarded friction, the generalised Langevin equation (GLE), 
is well known for the Gaussian noise. Then, for a given noise covariance, it allows us to determine 
all the fluctuations which, on the other hand, is not possible for non-Gaussian noises like $\xi(t)$. 
This process is interesting due to its jumping structure, convergent variance 
and non-trivial distributions resulting from GLE driven by $\xi(t)$. In this paper, 
we discuss those distributions for two different memory kernels: exponential and $1/t$. 

The paper is organised as follows. In Sec.II, the autocorrelation function for 
the Ornstein-Uhlenbeck process with the multiplicative L\'evy noise is derived both 
for the stable and truncated distribution. Sec.III is devoted to GLE driven by 
that process for the Cauchy distribution: 
the probability density distributions are simulated and the resulting fluctuations are 
compared with general analytical predictions. A similar analysis is performed for the case of 
the power-law covariance. Results are summarised in Sec.IV.

\section{Autocorrelation function for the multiplicative Ornstein-Uhlenbeck process} 

Dynamics of a massless particle subjected to a stochastic force and a linear 
deterministic force is determined by the following Langevin equation 
\begin{equation}
\label{la}
\dot \xi(t)=-\gamma\xi(t)+G(\xi)\eta(t), 
\end{equation}
where the $\xi$-dependent noise intensity $G(\xi)$ accounts for a nonhomogeneous 
form of the stochastic activation. We assume that increments of the stochastic force, 
$\eta(t)$, possess the stable and symmetric L\'evy distribution defined by 
a characteristic function $\exp(-K^\alpha|k|^\alpha)$, 
where $\alpha$ ($0<\alpha\le2$) is a L\'evy index and $G(\xi)=|\xi|^{-\theta}$. 
The system (\ref{la}) resolves itself to the ordinary Ornstein-Uhlenbeck process 
if $\theta=0$ and $\alpha=2$. In the multiplicative case, we must settle  
the stochastic integral interpretation that is decisive for the existence of the second 
moment: in the It\^o interpretation the asymptotic distribution of $\xi$ is the same 
as for $\eta$ whereas in the Stratonovich one the multiplicative factor essentially 
modifies the tail \cite{sro1}. The latter interpretation applies when the white noise 
is regarded as a limit of the correlated noise and the inertia is small \cite{sron}. 
Then the Langevin equation (\ref{la}) can be reduced to an equation with the 
additive noise by a simple change of the variable, 
\begin{equation}
\label{yodx}
y=\frac{1}{K(1+\theta)}|\xi|^{1+\theta}\hbox {sgn}(\xi). 
\end{equation}
The Fokker-Planck equation in the new variable takes the form 
\begin{equation}
\label{fraces}
\frac{\partial}{\partial t}p(y,t)=\gamma(1+\theta)\frac{\partial}{\partial y}
[yp(y,t)]+\frac{\partial^\alpha}{\partial |y|^\alpha}p(y,t)  
\end{equation}
and its solution, after transformation to the original variable, reads  
\begin{eqnarray} 
\label{soln}
p(\xi,t)=\frac{1+\theta}{\alpha|\xi|}H_{2,2}^{1,1}\left[\frac{|\xi|^{1+\theta}}
{K(1+\theta)\sigma(t)^{1/\alpha}}
\left|\begin{array}{l}
(1,1/\alpha),(1,1/2)\\
\\
(1,1),(1,1/2)
\end{array}\right.\right], 
\end{eqnarray}
where 
\begin{equation}
\label{as}
\sigma(t)=\frac{1-\exp[-\gamma\alpha(1+\theta) t]}{\gamma\alpha(1+\theta)} 
\end{equation}
and the initial condition $p(\xi,0)=\delta(\xi)$ has been assumed. Asymptotics of Eq.(\ref{soln}) 
is a power-law: $p(\xi,t)\sim|\xi|^{-1-\alpha-\alpha\theta}~~(|\xi|\gg1)$. 
In the limit $t\to\infty$, the system reaches a stationary state which is characterised by the variance 
\begin{equation}
\label{varxi}
\langle \xi^2\rangle=
-\frac{2}{\pi}K^{2/(1+\theta)}\alpha^{-1-2/\alpha(1+\theta)}(1+\theta)^{(2\alpha-2)/\alpha(1+\theta)}
\gamma^{-2/\alpha(1+\theta)}
\Gamma\left(-\frac{2/\alpha}{1+\theta}\right)
\Gamma\left(1+\frac{2}{1+\theta}\right)\sin\left(\frac{\pi}{1+\theta}\right) 
\end{equation}
if $\theta>2/\alpha-1$ ($\alpha<2$). 

For $\alpha=2$, a formal expression for the correlation function 
can be derived by means of an expansion of the general Fokker-Planck equation solution 
into its eigenfunctions \cite{schen}. The asymptotic 
behaviour appears exponential and the rate is given by the lowest eigenvalue. However, 
this conclusion may be wrong if a continuous spectrum is not negligible. This happens, for example, 
for the linear problem ($\theta=-1$) and then the exponential is modified by an algebraic 
term \cite{gra}. The system (\ref{la}) for $\alpha<2$ and with $\theta=0$ 
has a finite relaxation time but its quantification in terms of the covariance function 
is possible only after either a modification of this quantity \cite{samr} or by introducing 
a cut-off in the distribution. In the case of the multiplicative noise, for $\theta>2/\alpha-1$, 
the autocorrelation function exists and can be expressed by the integral 
\begin{equation}
\label{defc}
{\cal C}(t)=\langle\xi(0)\xi(t)\rangle=
\int\int\xi_1\xi_2p(\xi_2,t;\xi_1,0)d\xi_1d\xi_2=\int\int\xi_1\xi_2p(\xi_2,t|\xi_1,0)p(\xi_1)d\xi_1d\xi_2
\end{equation}
where $p(\xi)=\lim_{t\to\infty}p(\xi,t)$. In terms of the transformed variables, 
$C(t)$ assumes the following form
\begin{equation}
\label{defc1}
{\cal C}(t)=[K(1+\theta)]^{\frac{2}{1+\theta}}
\int\int |y_1|^{\frac{1}{1+\theta}}|y_2|^{\frac{1}{1+\theta}}
\hbox {sgn}(y_1)\hbox {sgn}(y_2)p(y_2,t|y_1,0)p(y_1)dy_1dy_2. 
\end{equation}
The conditional probability in Eq.(\ref{defc1}) is given by 
\begin{eqnarray}
\label{conpr}
p(y_2,t|y_1,0)=
\frac{1}{K\alpha\sigma(t)^{1/\alpha}}H_{2,2}^{1,1}\left[\frac{1}{K\sigma(t)^{1/\alpha}}
|y_2-y_1\hbox{e}^{-\gamma_\theta t}|
\left|\begin{array}{l}
(1-1/\alpha,1/\alpha),(1/2,1/2)\\
\\
(0,1),(1/2,1/2)
\end{array}\right.\right], 
\end{eqnarray}
where $\gamma_\theta=\gamma(1+\theta)$, whereas 
\begin{eqnarray}
\label{pody1}
p(y_1)=
\frac{(\alpha\gamma_\theta)^{1/\alpha}}{K\alpha}H_{2,2}^{1,1}\left[\frac{(\alpha\gamma_\theta)^{1/\alpha}}{K}
|y_1|\left|\begin{array}{l}
(1-1/\alpha,1/\alpha),(1/2,1/2)\\
\\
(0,1),(1/2,1/2)
\end{array}\right.\right]. 
\end{eqnarray}

The integral (\ref{defc1}) can be estimated in the long-time limit; 
details of the derivation are presented in Appendix A. The final expression reads 
\begin{equation}
\label{cdu}
{\cal C}(t)\approx 4K^\frac{3}{1+\theta}(1+\theta)^{\frac{2}{1+\theta}}\alpha^{\frac{1}{\alpha}
\frac{\theta}{1+\theta}-1}\gamma_\theta^{-\frac{1}{\alpha}\frac{1}{1+\theta}}
\frac{\Gamma(\frac{\theta/\alpha}{1+\theta})}{\Gamma(\frac{\theta/2}{1+\theta})}I\hbox{e}^{-\gamma_\theta t},
\end{equation}
where $I$ is given by Eq.(\ref{A.4}). However, the expansion (\ref{A.1}) contains infinite terms if 
$\alpha<2$; in particular, the asymptotic expansion of the Fox function in Eq.(\ref{A.4}) 
produces integrand tail of the form 
$y_1^{1/(1+\theta)-\alpha}$ and then $I$ diverges for any $\theta$ if $\alpha\le1$. 
Therefore, the approximation (\ref{cdu}) is not valid for the general stable distributions. 
However, one can argue that in many systems very long jumps do not 
emerge and it is reasonable to introduce 
a truncation of the distribution. Such a truncation can be realised as a simple cut-off or 
by inserting a fast-falling tail; typical forms are the exponential \cite{kop} and 
a power-law \cite{sok,che1}. Systems involving the multiplicative L\'evy noise were considered 
from that point of view in Ref.\cite{physa}. Convergence to the normal distribution, 
expected in this case, is so slow that 
it is not observed in the numerical simulations. The cut-off at some value of $|\eta|$ implies 
a finite upper integration limit in Eq.(\ref{A.4}) and $I$ becomes convergent.
Then the autocorrelation function falls exponentially with time; the rate does not depend on 
$\alpha$ and rises with $\theta$. The above result is valid also for the additive noise, $\theta=0$. 

The following analysis is restricted to the Cauchy distribution of the noise $\eta(t)$ ($\alpha=1$). 
The integral (\ref{defc1}) for the case without the truncation has been evaluated numerically. 
Inserting the conditional probability 
\begin{equation}
\label{conprc}
p(y_2,t|y_1,0)=\frac{K^2}{\pi\gamma_\theta}\frac{1-\exp(-\gamma_\theta t)}
{(y_2-y_1\exp(-\gamma_\theta t))^2+K^4(1-\exp(-\gamma_\theta t))^2/\gamma_\theta^2}
\end{equation}
to Eq.(\ref{defc1}) yields the expression for ${\cal C}(t)$; it is presented 
in Fig.1 for some values of $\theta$. 
The figure reveals a stretched exponential shape, $\exp(-\lambda t^\beta)$, and 
the parameter $\beta$ rises monotonically from 1.040 for $\theta=1.5$ to 1.081 for $\theta=5$. 
Since $\beta$ is close to 1, deviation from the simple exponential emerge only 
for very small values of ${\cal C}(t)$ (large $t$) and/or large $\theta$. Therefore, ${\cal C}(t)$ 
can be reasonable approximated by the dependence 
\begin{equation}
\label{cex}
{\cal C}(t)=\frac{(K/\gamma)^{2/(1+\theta)}}{\cos(\pi/(1+\theta))}\hbox{e}^{-\lambda t}, 
\end{equation}
where ${\cal C}(0)$ follows from Eq.(\ref{varxi}) and $\lambda$ is a parameter. 
Results for the truncated distributions, also presented in the figure, exhibit 
the fast-falling exponential tail, in agreement with Eq.(\ref{cdu}), 
and they coincide with the stable case at small $t$. 
\begin{center}
\begin{figure}
\includegraphics[width=12cm]{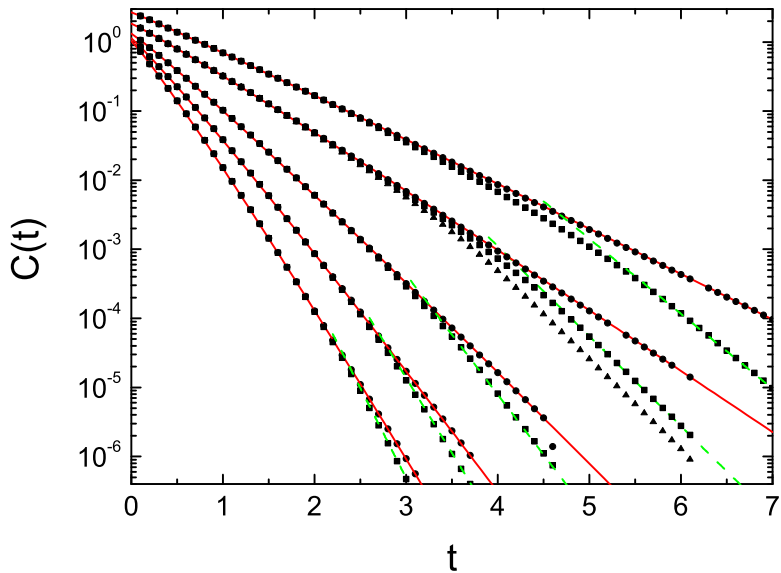}
\caption{(Colour online) The autocorrelation function of $\xi(t)$ for $\alpha=1$ 
and $\gamma=1$, calculated 
from Eq.(\ref{defc1}), for the following values of $\theta$: 1.5, 2, 3, 4, and 5 (points, 
from right to left). Results for the distribution truncated at $10^6$ and $10^5$ 
are marked by squares and triangles, respectively. The red solid line represents the stretched 
exponential function and the dashed green line has the shape $\hbox{e}^{-\gamma_\theta t}$.}
\end{figure}
\end{center}
By rescaling the noise $\eta(t)$ we get rid of $K$ and then $\lambda$ 
is completely determined by $\theta$ and $\gamma$. 
Fig.2 demonstrates that dependence on both parameters is simple; the expression for 
$\lambda(\theta,\gamma)$ can be uniquely determined from those results: 
\begin{equation}
\label{lamtg}
\lambda=0.80(\theta+0.31)\gamma\equiv c_\theta\gamma.
\end{equation}
The formula (\ref{cex}) with $\lambda$ from Eq.(\ref{lamtg}) and $\langle \xi^2\rangle$ 
from Eq.(\ref{varxi}) will be applied in Sec.III as the approximation to Eq.(\ref{defc1}) 
if time is not very large. 

\begin{center}
\begin{figure}
\includegraphics[width=12cm]{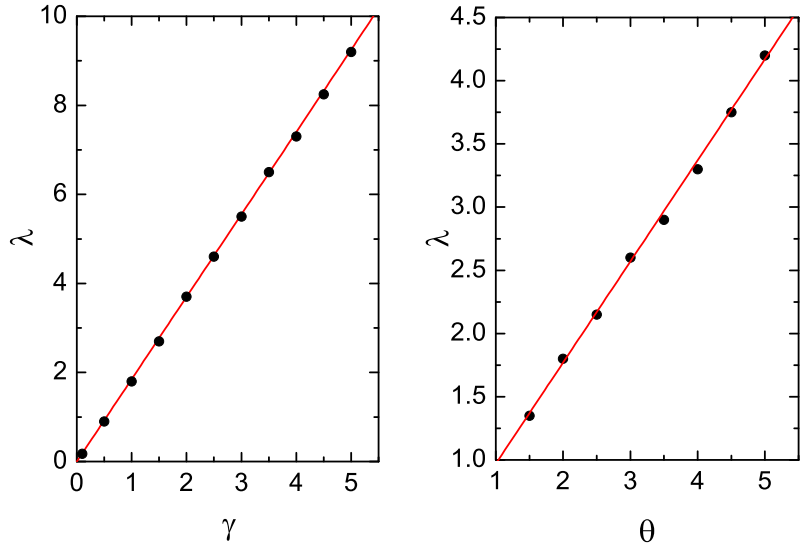}
\caption{(Colour online) Dependence $\lambda(\gamma)$ and $\lambda(\theta)$ (points). 
The straight lines mark the functions $1.85\gamma$ and $0.17+0.8\theta$ for the left 
and right panel, respectively.}
\end{figure}
\end{center}

\section{Memory effects in the dynamics} 

When the dynamics proceeds in a medium of a nonhomogeneous structure, one can expect 
nonlinear effects and non-Gaussian distributions. For example, a Langevin description for the case 
of a Brownian particle interacting with a general non-Gaussian thermal bath 
resolves itself to a nonlinear, multiplicative Langevin equation 
with a non-Gaussian white noise and nonlinear friction term which follows from the detailed 
balance symmetry \cite{dubnl}. If the equilibrium state of a stochastic system results 
from an interplay between an internal noise and damping, the noise intensity and the dissipation 
have to be mutually related (the Einstein relation). For a correlated noise and a linear coupling 
in the thermal bath, that relation requires 
a retarded friction in the Langevin equation which then becomes a linear integro-differential 
equation \cite{mori,lee}. Memory effects are important also for processes involving 
the L\'evy stable noise. It has been demonstrated in Ref. \cite{sron} that 
the external noise relaxation time modifies a slope of the power-law density distribution. 
The memory makes the friction term nonlocal in time. The fractional Langevin 
equation was introduced by Lutz \cite{lut} for the Gaussian noise which is distinguished due to 
the central limit theorem. For more general cases, e.g. in complex systems, an ordinary 
central limit theorem is no longer valid and the effective random force may assume a form different 
from the Gaussian even if a coupling within the thermostat is linear. 
GLE may be applied to such non-Gaussian processes \cite{cof} but in this case 
higher moments cannot be expressed by the first and second moments. 
We assume that dynamics is governed by a Langevin equation with the retarded friction and 
driven by the effective random force $\xi(t)$, defined by Eq.(\ref{la}). It satisfies 
the second fluctuation-dissipation theorem (FDT) \cite{kubo} to ensure a proper thermal equilibrium. 
Such a description is possible since FDT requires the existence of only first and second moments. 
Then we consider GLE in the form 
\begin{equation}
\label{gle}
m \frac{dv(t)}{dt} = -m\int_0^t K(t-\tau)v(\tau)d\tau + \xi(t), 
\end{equation}
where $v(t)$ is a velocity. FDT implies that the memory kernel has the same form as the noise covariance, 
$K(t)={\cal C}(t)/mT$, where $T$ is the temperature and the Boltzmann constant is set at one.
The equipartition energy rule is satisfied: 
$\langle v(\infty)^2\rangle=T/m$. Applying the Laplace transformation yields the solution, 
\begin{equation}
\label{solv}
v(t)=R(0)v_0+m^{-1}\int_0^t R(t-\tau)\xi(\tau)d\tau,
\end{equation}
with the initial condition $v(0)=v_0$, where the Laplace transform of 
the resolvent $R(t)$ is given by the equation 
\begin{equation}
\label{rods}
\widetilde R(s)=1/[s+\widetilde K(s)]. 
\end{equation}
All the fluctuations, if they exist, are determined by the resolvent $R(t)$. 
The energy equipartition rule follows from Eq.(\ref{solv}) in the limit $t\to\infty$. 
The resolvent $R(t)$ has an interpretation of the velocity autocorrelation function, 
${\cal C}_v=\langle v_0v(t)\rangle=(T/m)R(t)$, and it determines a speed of the relaxation to 
the equilibrium \cite{adel}: 
\begin{equation}
\label{relax}
\langle[v(t)-R(t)v_0]^2\rangle=\frac{T}{m}[1-R^2(t)]. 
\end{equation}
\begin{center}
\begin{figure}
\includegraphics[width=12cm]{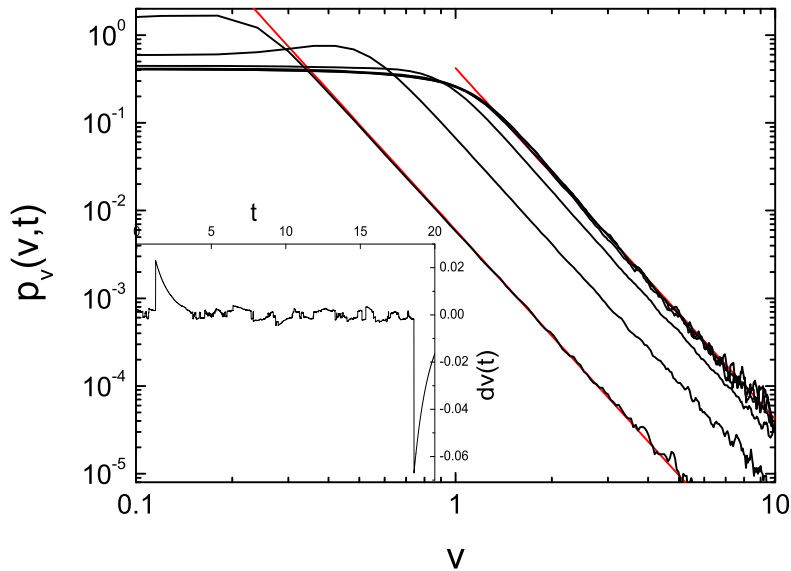}
\caption{(Colour online) The velocity distribution for $\theta=2$ and $\gamma=1$ 
at $t=0.2$, 0.5, 1, 2, 5, and 10 (from left to right), calculated from GLE, Eq.(\ref{gle}). 
The straight red lines mark the dependence $v^{-4}$. Inset: velocity increments 
for a single trajectory.}
\end{figure}
\end{center}

We assume that the driving noise $\xi(t)$ is given by Eq.(\ref{la}) and approximate 
its autocorrelation function by the exponential dependence (\ref{cex}). Then a straightforward 
calculation yields 
\begin{eqnarray}
\label{rodt}
R(t)=
\begin{cases} 
{1/(2\sqrt{-\Delta})
\left(B \hbox{e}^{-At}-A\hbox{e}^{-Bt}\right)~~~~~~~~~~~~~~~(\Delta <0)}\cr
{\hbox{e}^{-\lambda t/2}/\sqrt{\Delta}\left(\frac{\lambda}{2}
\sin\sqrt{\Delta}t+\sqrt{\Delta}\cos\sqrt{\Delta}t\right)~~(\Delta>0)}
\end{cases}
\end{eqnarray}
where $A=\lambda/2 -\sqrt{-\Delta}$, $B=\lambda/2 +\sqrt{-\Delta}$ and
$\Delta =\langle\xi^2\rangle/(mT)-\lambda^2/4$. 
One could expect that the probability density distribution, $p_v(v,t)$, 
converges to the normal distribution due to the finite variance. 
According to Eq.(\ref{solv}), the velocity is 
a linear combination of the weighted values of the noise, $R(t-t_i)\xi(t_i)\Delta t$, 
where $\Delta t$ is a constant integration step. The subsequent components are not independent 
but, since the autocorrelation ${\cal C}(t)$ falls with time, terms corresponding to 
times larger than some relaxation time of $\xi(t)$, $t_r\sim 1/\gamma$, 
can be regarded as independent and assumptions of the central limit theorem are satisfied. 
More precisely, for a sufficiently large $t$ the sums $S_n=\sum_{i=0}^n R(t-i\Delta t-nt_r)\xi(i\Delta t+nt_r)$, 
where $n=[t_r/\Delta t]$, are independent stochastic variables of finite variance. 
The variable $\sum_{n=1}^{[t/t_r]+1} S_n$ may converge with $t$ to the normal distribution if 
both $t/t_r$ is large for a non-zero $R(t)$ and fluctuations of $S_n$ are small. 
The latter condition emphasises importance of higher moments of $\xi(t)$. 
According to the Berry-Ess\'een theorem \cite{fel}, a distribution of a sum of $m$ mutually 
independent variables differs from the Gaussian by 
$33\langle \xi^3\rangle/(4\sigma^3\sqrt{m})$, where $\sigma$ is the standard deviation, 
providing the third moment is finite. Therefore, convergence to the normal distribution 
is not ensured if $\theta\le2$ and even for a larger $\theta$ it may be very slow. Deviations 
from the Gaussian are especially pronounced for large values of $|v|$ which correspond to jumps and 
such events are usually a result of single stochastic activations. In this case 
a distribution is similar to $p(\xi)$ and tails have the form $|v|^{-2-\theta}$.
On the other hand, events that produce small $|v|$ correspond to 
the trajectories consisting of many small segments, similar to the ordinary Brownian motion,  
and fluctuations are small; then we may expect the Gaussian shape. 

The Monte-Carlo simulations confirm presence of the power-law tails. Time evolution of $p_v(v,t)$, 
presented in Fig.3 for the case of the infinite third moment, indicates no trace 
of a convergence with time to the Gaussian; the distribution apparently reaches a stationary state 
near $t=2$ and the power-law dependence, $|v|^{-2-\theta}$, dominates the distribution. 
Trajectories reveal a jumping structure typical for the L\'evy flights. This structure 
is clearly visible when we plot the velocity increments for the discretized integral 
(\ref{solv}) (Fig.3). Note the finite jump relaxation time. 

Speed of the equilibration is governed by the parameter $A$ in Eq.(\ref{rodt}). 
Dependence on the parameters follows from Eq.(\ref{varxi}),(\ref{lamtg}); an estimation for large $\gamma$ 
yields $A\sim\gamma^{-2/(1+\theta)-1}/mT$. We conclude that the equilibration time ($\sim1/A$) 
rises with $\gamma$ because the noise intensity declines. 
Fig.4 presents equilibration of the variance for different sets of the parameters. The equilibrium value, 
$\langle v^2\rangle=T/m$, is reached at short time for small both $\gamma$ and $\theta$ since then 
the noise intensity is large. Differences between results of the Monte Carlo calculations and 
Eq.(\ref{relax}) are due to the approximation of the exact ${\cal C}(t)$ by the exponential; 
the equilibrium value is slightly overestimated compared to the energy equipartition rule. 
\begin{center}
\begin{figure}
\includegraphics[width=12cm]{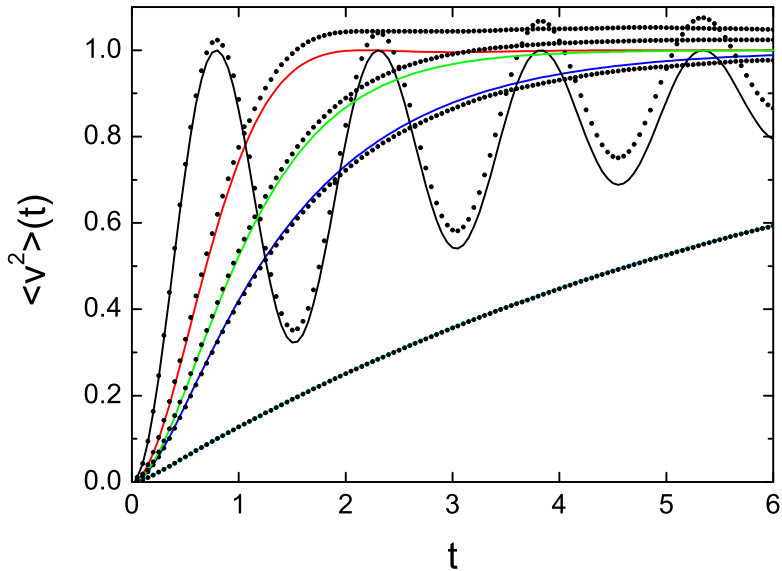}
\caption{(Colour online) Time evolution of the variance calculated from the Monte Carlo 
simulations (points) and from Eq.(\ref{relax}) (lines) for $T=1$ and $m=1$ with the initial 
condition $v_0=0$. The following cases are presented 
(from left to right): 1. $\theta=3$ and $\gamma=0.1$ (black), 
2. $\theta=2$ and $\gamma=1$ (red), 3. $\theta=3$ and $\gamma=1$ (green), 
4. $\theta=2$ and $\gamma=2$ (blue), and 5. $\theta=2$ and $\gamma=5$ (cyan).}
\end{figure}
\end{center}
\begin{center}
\begin{figure}
\includegraphics[width=12cm]{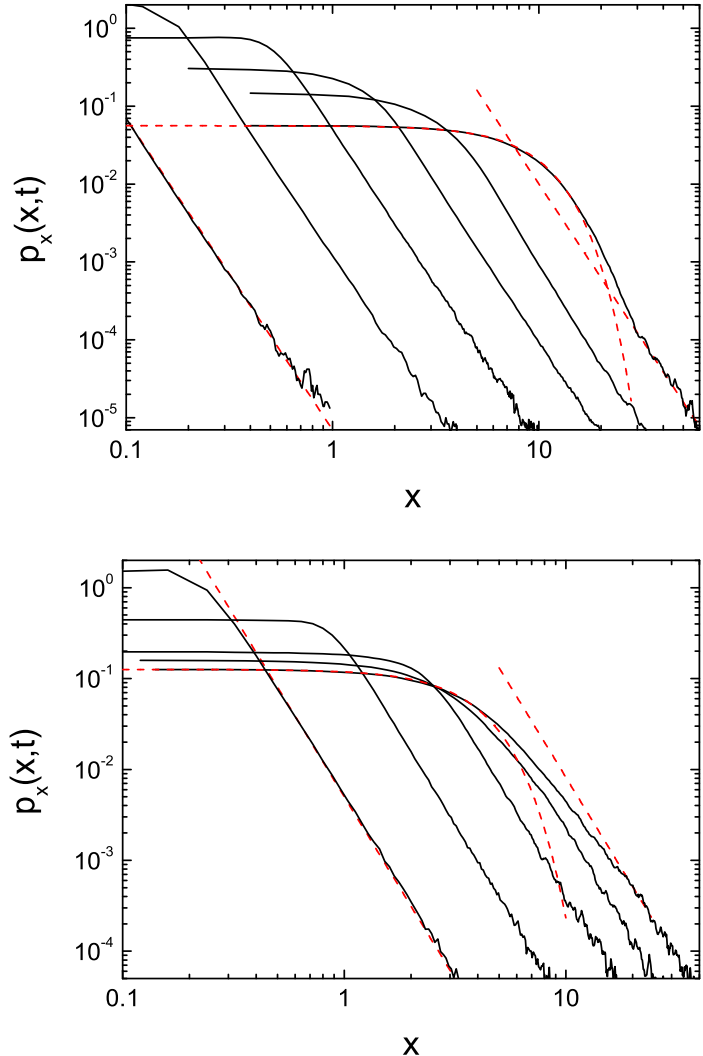}
\caption{(Colour online) Upper part: the position distribution for $\theta=2$ and $\gamma=1$ 
at $t=0.2$, 0.5, 1, 2, 5, and 30 (from left to right). The red dashed lines 
mark the dependence $x^{-4}$ and the Gaussian. Lower part: the same but for the covariance 
(\ref{1nadt}) and the following times: 1, 2, 5, 20 and 50.
}
\end{figure}
\end{center}

The position $x$ is given by an expression similar to Eq.(\ref{solv}) but with the resolvent 
$R_x(t)=\int_0^tR(t')dt'$. The distribution of $x$, $p_x(x,t)$, is presented in Fig.5 
as a function of time; we assumed the initial condition $x(0)=0$. Two limits, discussed above, 
are clearly visible if time is sufficiently long: $p_x(x,t)$ assumes the Gaussian shape 
for $|x|<20$ $(t=30)$ whereas the tail 
is of the form $|x|^{-2-\theta}$. The position variance directly follows from the identity 
\begin{equation}
\label{grku}
\langle x^2\rangle(t)=2\int_0^t(t-t'){\cal C}_v(t')dt', 
\end{equation}
where the averaging is performed over the equilibrium state. 
A direct evaluation for the case $\Delta>0$ yields 
\begin{equation}
\label{x2odt}
\langle x^2\rangle(t)=\frac{2mT^3}{\langle\xi^2\rangle^2}\left[\frac{\langle\xi^2\rangle}{mT}
-\lambda^2+\frac{\lambda\langle\xi^2\rangle}{mT}t+\hbox{e}^{-\lambda t/2}
\left((\lambda^2-\frac{\langle\xi^2\rangle}{mT})\cos(\sqrt{\Delta}t)+\frac{\lambda}{2\sqrt{\Delta}}
(\frac{\lambda^2}{4}-3\frac{\langle\xi^2\rangle}{mT})\sin(\sqrt{\Delta}t)\right)\right].
\end{equation}
We omit the analogous expression for $\Delta<0$. In the limit of large time 
the variance rises linearly with time and the diffusion 
coefficient ${\cal D}=\lim_{t\to\infty}\langle x^2(t)\rangle/2t=
T^2\lambda/\langle\xi^2\rangle=T^2c_\theta\cos(\pi/(1+\theta))\gamma^{1+2/(1+\theta)}$. 
The position variance as a function of time is presented in Fig.6. 
The Monte Carlo results, obtained by integration of Eq.(\ref{solv}), 
reveal a slightly stronger time-dependence than the linear growth predicted by Eq.(\ref{grku}): 
$t^\beta$ with $\beta=1.02, 1.04$ and 1.05 for $\theta=1.5$, 2 and 3, respectively. 
\begin{center}
\begin{figure}
\includegraphics[width=12cm]{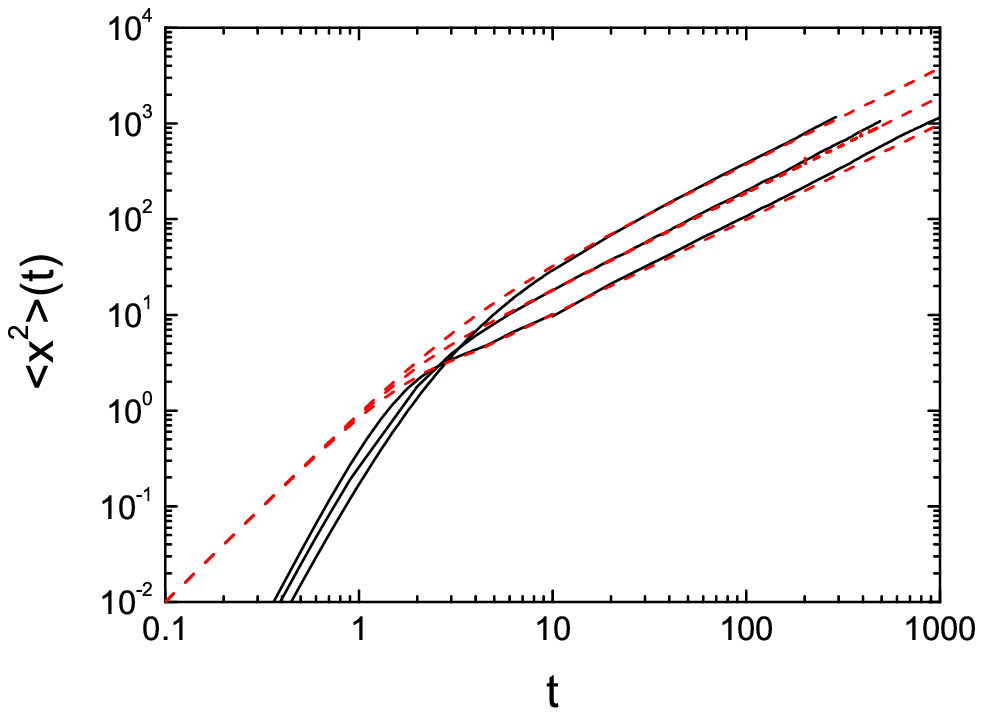}
\caption{(Colour online) Position variance obtained from integration of Eq.(\ref{solv}) 
with the initial condition $v_0=x(0)=0$ for $\gamma=1$ and $\theta=1.5, 2, 3$ 
(solid lines from bottom to top). Results marked by dashed red lines follow from Eq.(\ref{grku}).}
\end{figure}
\end{center}

In many physical problems the observed covariance functions are not exponential. 
The power-law form of the memory function was discussed 
in connection with a frictional resistance \cite{bouss} and in 
the hydrodynamics \cite{mazo}. In particular, diffusion in the dense liquids requires 
the memory function falling like $t^{-3/2}$ \cite{opp}. GLE for systems with power-law 
kernels, $|t|^{-\beta}~(\beta\ne1)$, takes a form of the fractional Langevin equation, 
where the damping term is expressed by the Riemann-Louville operator, 
$_0D_t^{\beta-1}$ \cite{lut}; power-law kernels are present in the fractional Brownian 
motion theory \cite{deng}. Long-time correlations are observed in the complex systems 
that usually possess non-Gaussian distributions with power-law tails. A very slow falling 
covariance, corresponding to a $1/f$ noise, was found in an analysis of absolute returns 
in the US market \cite{gvoz}. The autocorrelation function 
of the displacement may even rise with time; 
this effect was experimentally demonstrated for the diffusion in a dusty plasma liquids for which 
the corresponding probability distributions exhibit fat tails \cite{wai}. 
The covariance $1/t$, in turn, was observed in connection with the noise-induced Stark 
broadening \cite{fri} and obtained, for a two-dimensional system, from the Navier-Stokes
equations \cite{alder}. The presence of this form of the autocorrelation function may be related 
to a specific topology of the medium: it emerges when the trajectory has a structure 
of long straight-line intervals, like for the Lorentz gas \cite{gei}, and may be 
encountered in the nuclear reactions \cite{sropl}. In this paper, we solve GLE with the memory 
function in the form $1/t$. More precisely, we assume 
\begin{equation}
\label{1nadt}
{\cal C}(t)=(1-\hbox{e}^{-Lt})/Lt, 
\end{equation}
where $L=$const$>0$. 

According to the results of Sec. II, any process $\xi(t)$, given by Eq.(\ref{la}), is characterised by 
the exponential covariance and the rate is uniquely determined by $\lambda$. Assuming that 
$\xi(t)$ is an elementary process $\xi_\lambda(t)$, we can construct a compound process by 
a superposition of $\xi_\lambda(t)$ where the parameter $\lambda$ is regarded as 
a stochastic variable. Therefore, we may obtain an arbitrary, {\it a priori} assumed 
covariance by averaging, with a weight $\psi(\lambda)$, over an ensemble of trajectories 
corresponding to a fixed value of $\theta$ and different values of $\gamma$. Since, for a given $\theta$, 
${\cal C}(t)=\int_0^\infty \langle \xi^2\rangle \psi(\lambda)\hbox{e}^{-\lambda t}d\lambda\sim 
\int_0^\infty \lambda^{-2/(1+\theta)}\psi(\lambda)\hbox{e}^{-\lambda t}d\lambda$, the distribution 
of $\lambda$ can be evaluated for any ${\cal C}(t)$ by inversion of the Laplace transform: 
\begin{equation}
\label{cpodt}
\psi(\lambda)\sim \lambda^{2/(1+\theta)}{\cal L}^{-1}[{\cal C}(t)]. 
\end{equation}
Eq.(\ref{1nadt}) corresponds to the following normalised distribution 
\begin{equation}
\label{psi}
\psi(\lambda)=\frac{L^{2/(1+\theta)}}{2/(1+\theta)+1}c_\theta^{-2/(1+\theta)}\cos\frac{\pi}{1+\theta}
\lambda^{2/(1+\theta)}
\end{equation}
for $\lambda\in(0,L)$ and 0 elsewhere. The distribution of all the elementary processes 
$\xi_\lambda(t)$ has the same asymptotic form, $|\xi|^{-2-\theta}$, with the same slope 
of the tails since $\theta$ is fixed in the statistical ensamble. 
$\gamma$, in turn, influences a relative intensity of the noises $\xi_\lambda(t)$. 
Solution of GLE is given by Eq.(\ref{solv}) where transform 
of the memory kernel, $\widetilde K(s)=[\ln(s/L+1)-\ln(s/L)]/L$, follows from Eq.(\ref{rods}). 
Eq.(\ref{grku}) yields ${\cal D}=\lim_{t\to\infty}d\langle x^2(t)\rangle/dt=
\int_0^\infty{\cal C}_v(t)dt\sim {\widetilde R}(0)=0$ for any $L$. 
Inversion of the transform yields 
\begin{equation}
\label{rpot}
R(t)=\hbox{e}^{at}\left(c_1\cos(bt)+c_2\sin(bt)\right)-L\int_0^L\frac{\hbox{e}^{-tx}dx}
{[Lx-\ln(L/x-1)]^2+\pi^2}. 
\end{equation}
Details of the derivation and values of the coefficients, as well as some remarks about the numerics, 
are presented in Appendix B. 

The shape of the stationary velocity distribution for the covariance (\ref{1nadt}) is similar to that 
for the exponential covariance case but the dependence $1/|v|^{2+\theta}$ 
of the tails for $\theta=2$ shifts to the relatively large $|v|$ and the equilibration 
time is larger. The damping parameter $a$, given by 
Eq.(\ref{B.1}), non-monotonically depends on $L$ but it becomes very small for large $L$ and 
the time needed to reach the stationary state is then extremely long. 
The parameter $\theta$ does not influence the equilibration time but strongly modifies 
the distribution tail. For example, the tail assumes the shape $1/v^8$ for $\theta=4$, 
i.e. it falls stronger than the noise distribution. Anyway, a convergence to normal distribution 
is not observed. 

\begin{center}
\begin{figure}
\includegraphics[width=12cm]{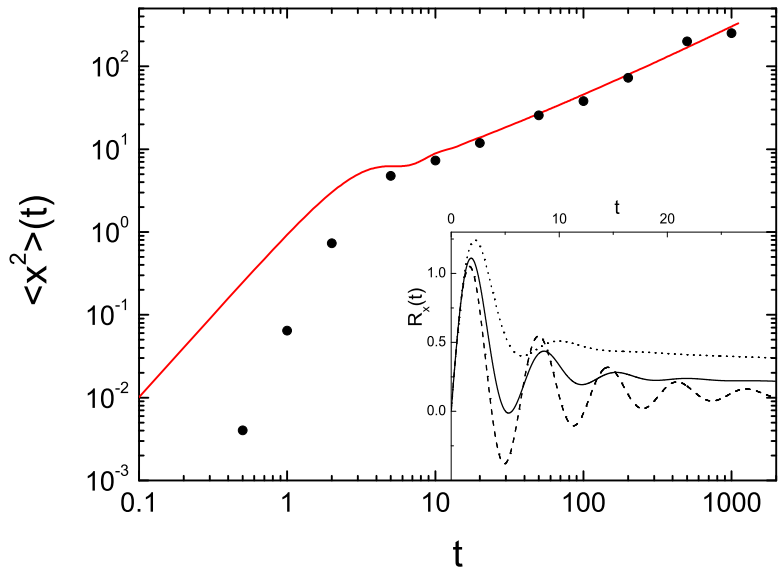}
\caption{(Colour online) Position variance obtained from integration of Eq.(\ref{solv}) 
with the initial condition $v_0=x(0)=0$ for the covariance (\ref{1nadt}), calculated 
with the parameters $L=1$ and $\theta=2$ (points). Results marked by the red line follow 
from Eq.(\ref{grku}). Inset: the resolvent $R_x(t)$ for 
$L=0.5$ (dashed line), 1 (solid line) and 2 (dotted line).}
\end{figure}
\end{center}
Distribution of the position was calculated by means of the integrated resolvent $R_x(t)$ and 
results are presented in Fig.5. The $x$-dependence of $p_x(x,t)$, in particular the asymptotics, 
is similar to the case of the exponential covariance. The main difference consist in 
the expansion speed: the distribution for the covariance (\ref{1nadt}) widens much slower. 
Time-dependence of the variance is given by Eq.(\ref{grku}), where the integral 
in (\ref{rpot}) has to be estimated numerically. The shape of the curve, shown in Fig.7, 
reveals an apparent shape $t^{0.8}$ at the long time which indicate the sublinear behaviour. 
As expected from the equation ${\widetilde R}(s=0)=0$, the system 
is subdiffusive and simulations agree with Eq.(\ref{grku}) in the stationary limit. 
However, asymptotics of the variance is in fact not a power-law. 
According to a conjecture in Ref. \cite{fer}, 
the position variance should behave in the limit of long time like $t/\ln(t)$ the form of which 
has been interpreted by the authors as an analogy to critical exponents in a phase transition.
One obtains a similar dependence when ${\cal D}(t)$ is estimated 
by establishing lower and upper limits of the integral \cite{kam}. The expression 
$\langle x^2\rangle(t)=0.91t/\ln(t)$ agrees with the exact result, Eq.(\ref{grku}), for $t>30$. 
The time-dependent diffusion coefficient is given by 
the resolvent ${\cal D}(t)\sim R_x(t)$ and it is also presented in Fig.7. It appears 
very sensitive on $L$: oscillations, being strong for small $L$, vanish quickly if $L$ is large.

\section{Summary and conclusions}

The overdamped Langevin equation with the quadratic potential and the multiplicative 
L\'evy stable noise describes a process that comprises a jumping structure of trajectories 
and convergent moments: variance and covariance. We have demonstrated that the autocorrelation function 
${\cal C}(t)$ for the truncated distribution falls exponentially, in the limit 
of a long time, with the $\alpha$-independent rate $\gamma(1+\theta)$. 
Correlations were studied in detail for the Cauchy distribution. 
It has been found that for the stable case ${\cal C}(t)$ obey the stretched-exponential form but 
can be reasonable approximated by the simple exponential. The rate has been uniquely determined 
as a function of the system parameters: it rises linearly with $\gamma$ and $\theta$. Higher moments 
may also be convergent if one chooses a sufficiently large $\theta$. The exponential decay 
for the truncated case is faster than that for the stable distribution, a conclusion that 
emphasizes a role of very long jumps in preserving the memory in the system. One may construct 
a stochastic process characterised by an arbitrary form of the covariance by a superposition 
of trajectories with different $\gamma$, i.e. by assuming the 
parameter $\gamma$ as a stochastic variable. Moreover, one can reproduce an arbitrary slope of 
the distribution tail since that is governed solely by the parameter $\theta$. The above 
properties of the process $\xi(t)$ suggest its applicability to problems which require 
both fat tails and long correlation time. 

If a stochastic force that obeys the above properties is balanced by the damping force, 
the fluctuation-dissipation theorem and the equipartition energy rule are satisfied; 
then the process obeys GLE, a fact that is well-known for the Gaussian case. 
We applied GLE to the case for which the driving force is given by $\xi(t)$. 
The equation predicts tails of both velocity and position distribution, 
dominated by single jumps, of the same form as the driving noise. The central part of 
the distribution, in turn, results from many small stochastic activations and for $p_x(x,t)$ 
converges to the Gaussian, whereas the intermediate region assumes the fast-falling 
power-law. Similar distributions may be observed in many complex systems since they are characterised 
by a substantial memory and the thermal equilibration is accompanied by rare but spectacular events.  
Transport properties of the system described by GLE follow directly from the noise covariance; 
the position variance rises linearly for the exponential covariance and sublinearly for the covariance 
$\sim 1/t$. Numerical trajectory simulations involving the process $\xi(t)$ confirm that general result. 
Therefore, jumps and power-law tails of the distribution may coexist with the thermal equilibrium. 

\section*{APPENDIX A}

\setcounter{equation}{0}
\renewcommand{\theequation}{A\arabic{equation}} 

We derive the expression for ${\cal C}(t)$ in a limit of large $t$, Eq.(\ref{cdu}). First, 
we expand the conditional probability, Eq.(\ref{conpr}), in powers of 
$\epsilon=\hbox{e}^{-\gamma_\theta t}$ to the first order. Expansion of the first term 
in Eq.(\ref{conpr}) yields $(1-\epsilon^\alpha)^{-1/\alpha}=1+\epsilon^\alpha/\alpha+\dots$. 
The Fox function is given by the series, 
\begin{equation}
\label{A.1}
H[\frac{1}{K\sigma^{1/\alpha}}|y_2-y_1\epsilon|]=H[\frac{(\alpha\gamma_\theta)^{1/\alpha}}{K}|y_2|]+
\frac{\partial}{\partial\epsilon}H(\epsilon=0)\epsilon+\dots,
\end{equation}
where the coefficients are dropped. The derivative involves the Fox function 
of the higher order \cite{mat,sri}: 
\begin{eqnarray}
\label{A.2}
\frac{\partial H}{\partial\epsilon}=-\frac{y_1}{y_2-y_1\epsilon}
H_{3,3}^{1,2}\left[\frac{1}{K\sigma^{1/\alpha}}|y_2-y_1\epsilon|
\left|\begin{array}{l}
(0,1),(1-1/\alpha,1/\alpha),(1/2,1/2)\\
\\
(0,1),(1/2,1/2),(1,1)
\end{array}\right.\right]. 
\end{eqnarray}
Next, we insert the above expansions to Eq.(\ref{defc1}) and neglect terms of a higher order 
than $\epsilon$. The first component vanishes 
because the double integral can be factorised and both integrands are odd. The integral 
over $y_2$ resolves itself to a Mellin transform: 
\begin{equation}
\label{A.3}
\int_0^\infty y_2^{-\theta/(1+\theta)}H_{3,3}^{1,2}[\frac{(\alpha\gamma_\theta)^{1/\alpha}}{K}y_2]dy_2=
K^{1/(1+\theta)}(\alpha\gamma_\theta)^{-1/\alpha(1+\theta)}
\chi(-\frac{1}{1+\theta})=-K^{1/(1+\theta)}(\alpha\gamma_\theta)^{-1/\alpha(1+\theta)}
\frac{\Gamma(\frac{\theta/\alpha}{1+\theta})}{\Gamma(\frac{\theta/2}{1+\theta})}, 
\end{equation}
where $\chi$ stands for the Mellin transform from $H_{3,3}^{1,2}$. Elimination of the 
algebraic factor in the integral over $y_1$ yields 
\begin{eqnarray}
\label{A.4}
I=\int_0^\infty
H_{2,2}^{1,1}\left[x
\left|\begin{array}{l}
(1+\frac{1}{\alpha(1+\theta)},\frac{1}{\alpha}),(1+\frac{1}{2}\frac{1}{1+\theta},\frac{1}{2})\\
\\
(\frac{2+\theta}{1+\theta},1),(1+\frac{1}{2}\frac{1}{1+\theta},\frac{1}{2})
\end{array}\right.\right]dx. 
\end{eqnarray}

\section*{APPENDIX B}

\setcounter{equation}{0}
\renewcommand{\theequation}{B\arabic{equation}} 

We derive the expression (\ref{rpot}) where the integral $\int_{\sigma-i\infty}^{\sigma+i\infty} 
{\widetilde R}(z)\hbox{e}^{tz}dz$ is to be evaluated. The contour consist of a straight 
line parallel to the imaginary axis at $\sigma>0$, a large half-circle in the left half-plane and 
a cut along the real segment $(-L,0)$. Roots of the equation 
\begin{equation}
\label{B.1}
Lz+\ln(z/L+1)-\ln(z/L)=0 
\end{equation}
are of the form $z_{1,2}=a\pm bi$ and they have to be found 
numerically for a given $L$. After a straightforward evaluation of the sum 
over residues, we obtain the first component of Eq.(\ref{rpot}) where the coefficients are 
$c_1=2[a^4 + 2 a^3 L + a (2 b^2 - 1) L + a^2 (2 b^2 + L^2 - 1) + b^2 (b^2 + L^2 + 1)]/A$, 
$c_2=2 b (2 a + L)/A$ and 
$A=1+a^4 + b^4 + 2 a^3 L + 2 a (b^2 - 1) L + b^2 (2 + L^2) + 
a^2 (2 b^2 + L^2 - 2)$. Contribution from both branches along the cut resolves itself 
to the integral in Eq.(\ref{rpot}). 

Trajectory numerical simulations require a value of $R(t)$ for each integration step. 
Approximation of the asymptotics is easy to determine. For example, for $\theta=2$ and $L=1$ we get 
$R(t)=-0.113t^{-1.292}~~(t>60)$, a formula that coincides with the numerical integration up to 
at least $t=2000$. $R(t)$ for small $t$ was evaluated with a step 0.001 and stored.

\end{document}